\documentclass[twocolumn,showpacs,prl,10pt]{revtex4-1}
\usepackage[english]{babel}
\usepackage{amsmath,graphicx,amssymb} 
\usepackage{times}
\usepackage{epstopdf} 

\begin{document} 
\title{Unified description of hydrogen bonding
and proton transfer
by a two-state effective Hamiltonian }

\author{Ross H. McKenzie}
\email{email: r.mckenzie@uq.edu.au}
\affiliation{School of Mathematics and Physics, University of Queensland,
  Brisbane, 4072 Queensland, Australia} 
\date{\today}
                   
\begin{abstract}
  An effective Hamiltonian is considered which
describes hydrogen bonding
and proton tranfer
between two molecules due to the quantum mechanical
interaction between the orbitals of the H-atom and 
of the donor  (D) and acceptor (A) atoms in the molecules.
The Hamiltonian acts on two diabatic states and has a simple 
chemically motivated form for its matrix elements.
The model gives insight into the "H-bond puzzle", describes different
classes of bonds, and 
empirical correlations between the donor-acceptor distance $R$
and binding energies, D-H bond lengths, and the softening of
D-H vibrational frequencies.
A key prediction is 
the UV 
photo-dissociation of H-bonded complexes via an excited electronic state with
an exalted vibrational frequency.

\end{abstract}

\maketitle 

Hydrogen bonds play a key role    in            
a diverse range of phenomena in physics, chemistry, molecular biology,
and materials science. For example, 
hydrogen bonding is central to the unique
properties of water, to 
protein folding and function, to proton transport \cite{BerkelbachPRL09},
to corrosion \cite{LiPRL10}, and to aspects of crystal engineering.
Hydrogen bonds have properties that are distinctly different from
other chemical bonds. Many properties
are poorly understood and recently the International
Union of Pure and Applied Chemistry gave
a new definition \cite{DesirajuACIE}, particularly motivated
by the occurrence of hydrogen bonding to a wide range
of atoms other than oxygen (e.g., flourine, carbon, nitrogen)
and a wide range of bond lengths and energies.
 Gilli and Gilli have 
emphasized "the H-bond puzzle" which is stated as
"the unique feature of the H-bond is that bonds made by the same donor-acceptor pair may display an extremely wide range of
 energies and geometries" \cite{Gilli09}. 
A wide range of empirical correlations
have been observed between different
physical properties  including
bond lengths, binding energies, shifts in vibrational frequencies,
vibrational absorption intensity, 
and NMR chemical shifts \cite{SteinerACIE,Gilli09}.

\begin{figure}[htb] 
\centering 
\includegraphics[width=64mm]{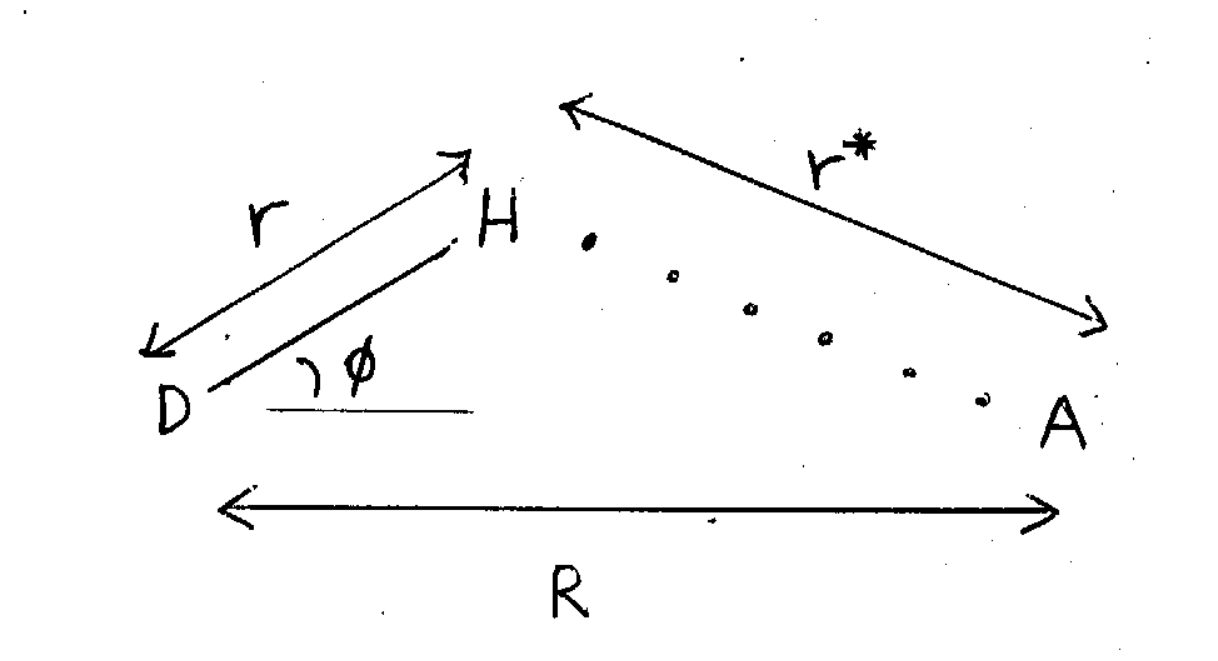}
\caption{(Color online.) 
Definition of geometric variables for a hydrogen bond between
a donor D and an acceptor A.
}
\label{fig1}
\end{figure}

In this Letter, I consider a simple effective
two state Hamiltonian
for a hydrogen bond between a donor denoted D-H and and an acceptor,
which form a complex, denoted D-H$\cdots$A. The Hamiltonian
provides insight into the H-bond puzzle,
gives a single description of different classes of H-bonds,  and
provides a semi-quantitative description of 
observed empirical correlations.
The key independent variable (or physio-chemical "descriptor") 
is the distance between the donor and acceptor $R$
 (see Figure \ref{fig1}).

{\it Reduced Hilbert space for the effective Hamiltonian.}
Diabatic states \cite{PacherJCP88}
(including valence bond states) have
proven to be a powerful method of developing
 chemical concepts \cite{Shaik,PolitzerJPCL10}.
Previously it has been proposed that hydrogen bonding
and hydrogen transfer reactions can be described by 
an Empirical Valence Bond model\cite{Horiuti,CoulsonAF54,Warshel,BorgisJMS97,SagnellaJCP98,FlorianJPCA02}
involving valence bond states.
Here, I choose a basis consisting
of  two diabatic states which can be denoted
as $|$D-H,A$>$ and $|$D,H-A$>$.
The latter represents a product state of the electronic states
of a D atom and of a H-A bond in the absence of
the D atom.
These  involve D-H and H-A bonds which have both covalent and ionic
components, the relative weight of which depends on the distance
$r$. The Morse potentials (see below) capture the associated energetics
including the partial electrostatic character of the H-bond.

{\it Effective Hamiltonian.}
The   Hamiltonian for the diabatic states
will have matrix elements which depend on
the D-H bond length $r$, the donor-acceptor separation $R$,
and the angle $\phi$ which describes the deviation from linearity 
(compare Figure \ref{fig1}).
The functional dependence can be parametrised 
from quantum chemistry calculations \cite{SagnellaJCP98}.
However, the parameters may depend significantly on the level
of theory used.
The main point of this Letter is that
one can obtain both a qualitative and semi-quantitative
description of hydrogen bonding using
a simple and physically transparent parametrisation
of these matrix elements.
This approach highlights the quantum mechanical (covalent)
character of the H-bond \cite{IsaacsPRL99} and
unifies H-bonding involving different atoms and
weak, medium, and strong (symmetrical) H-bonds.
The latter are sometimes characterised as covalent 4 electron,
3 centre bonds \cite{Shaik}.
The model Hamiltonian has the advantage that it is
straightforward to extend it to 
describe nuclear quantum effects (including going beyond the 
Born-Oppenheimer approximation),  multiple H-bonds, and 
collective effects.

The Morse potential describes the energy of a single bond 
within one of the molecules in the absence of the second 
(and thus the diabatic states).
The two cases $j=D,A$ denote the
donor D-H bond and acceptor A-H bond, respectively.
The Morse potential is
\begin{equation}
V_j(r) = D_j[\exp(-2a_j(r-r_{0j})) - 2\exp(-a_j(r-r_{0j}))]
\label{morse}
\end{equation}
where $D_j$ is the binding energy, $r_{0j}$ is the equilibrium bond length,
and $a_j$ is the decay constant. 
The harmonic vibrational frequency $\omega$ is given by
$\mu\omega^2=2D_j a_j^2$ where $\mu$ is the reduced mass.
[For O-H bonds, $\omega \simeq 3600 $ cm$^{-1}$, $D \simeq 120$ kcal/mol,
$a \simeq 2.2/\AA$, $r_0 \simeq 0.96 \AA$.]
A simple harmonic potential is not sufficient because
the O-H bond is highly anharmonic and we
will be interested in regimes where there is considerable
stretching of the bonds.

I  take the effective Hamiltonian describing the two interacting
diabatic states to have the form
\begin{equation}
H = \left(\begin{array}{cc} 
V_D(r)           & 
\Delta_{DA}(R,\phi) \\
\Delta_{DA}(R,\phi)  & V_A(r^*)             
\end{array}\right)
\end{equation}
where the diabatic states are coupled  via the matrix element
\begin{equation}
\Delta_{DA}(R,\phi  )=
\Delta_0 \cos(\phi) 
\frac{(R-r\cos\phi)}
{r^*}
 \exp(-b R)
\label{DeltaR}
\end{equation}
where $r^* = \sqrt{R^2+r^2-2rR\cos\phi}$ is the length of the A-H bond
 (see Figure \ref{fig1}),
and $b$ defines the decay of the matrix element with
increasing $R$.
This functional dependence on $R$ and $\phi$ can be justified from
that for orbital overlap integrals \cite{MullikenJCP49}
together with a valence bond theory description of 4 electron 3 orbital 
systems \cite{Shaik}.
With regard to the angular dependence
I have assumed that the $\sigma$ overlap dominates the $\pi$ overlap.
Roughly $\Delta_{DA}$ is the overlap of the hybrid (s and p) orbitals
on the D and A atoms and there will be some variation in
the parameters $\Delta_0$ and $b$ with the chemical
identity of the atoms D and A.  
For the rest of the paper I focus on the case of linear H bonds ($\phi=0$)
and (\ref{DeltaR}) can be written as
\begin{equation}
\Delta(R)=\Delta_1 \exp(-b(R-R_1))
\label{deltaR}
\end{equation}
 where $R_1$ is a reference distance, $R_1 \equiv 2r_0+1/a$.

{\it Potential energy surfaces.}
In the adiabatic limit the energy eigenvalues are
\begin{eqnarray}
E_\pm(r,R) &=& 
\frac{1}{2}(V_D(r) + V_A(R-r)) \\
&\pm& 
\frac{1}{2}\left((V_D(r) - V_A(R-r))^2 + 4\Delta(R)^2
\right)^{\frac{1}{2}}.
\end{eqnarray}
I  now focus on the symmetric case $D=A$ and
return to the asymmetric case briefly at the end of
the paper.

\begin{figure}[htb] 
\centering 
\includegraphics[width=64mm]{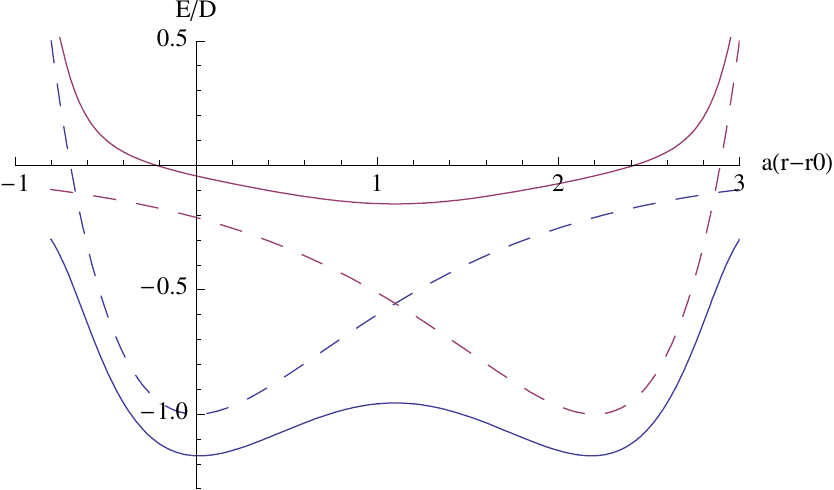}
\includegraphics[width=64mm]{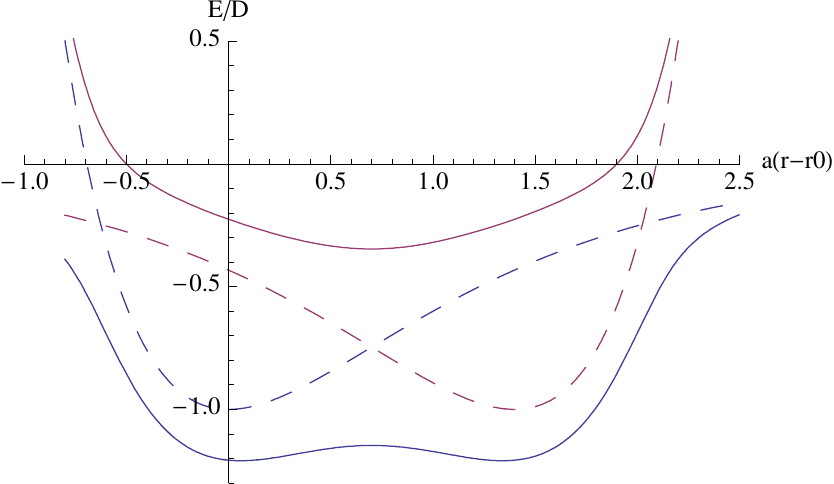}
\includegraphics[width=64mm]{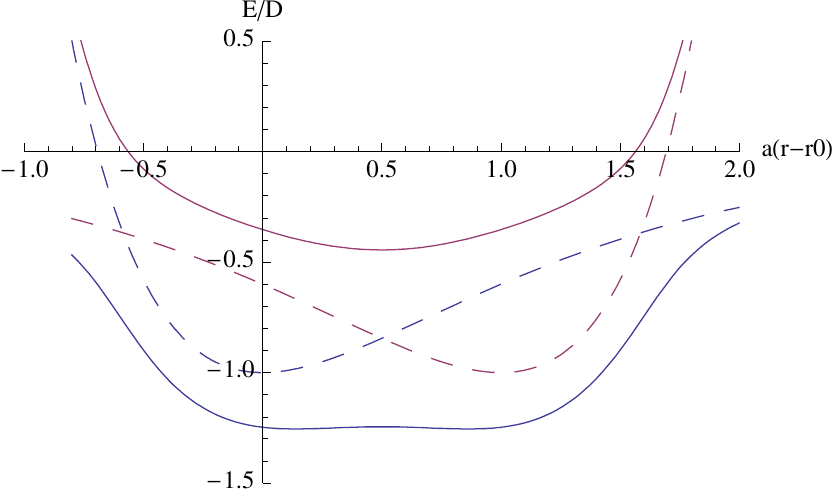}
\caption{(Color online.) 
Potential energy curves for the diabatic and adiabatic states
of a symmetric hydrogen bonded system.
The horizontal axis is proportional to the difference
between the length of the D-H bond, $r$,
and    its  isolated value, $r_0$.
The vertical energy scale is $D$, the binding energy of an isolated
D-H bond.
All adiabatic curves are for $\Delta_1 =0.4 D$ and $b=a$.
The diabatic curves (dashed lines) are Morse potentials
centred at $r=r_0$ and $r^*=R-r_0$ and correspond to
isolated D-H and H-A bonds, respectively.
For parameters relevant to a O-H$\cdots$O system
the three sets of  curves correspond (from top to bottom) 
to oxygen atom separations of
$R=2.9$, 2.6, and 2.3 $\AA$, respectively,
characteristic of weak, moderate (low barrier),
 and strong hydrogen bonds \cite{Gilli09}.
An important prediction of this model is the existence of the
excited state with energy $E_+(r,R)$ (upper curve).
}
\label{fig2}
\end{figure}

Figure \ref{fig2} shows 
 the adiabatic energy eigenvalues (potential energy curves)
$E_{-}(r,R)$
and
$E_+(r,R)$
as a function of $r$,  for three different fixed $R$ values.
Note three qualitatively different curves,
corresponding to weak, moderate, and strong hydrogen bonds.
For moderate bonds (corresponding to $R \simeq 2.6 \AA$ for an
O-H$\cdots$O system) the energy barrier becomes comparable to 
the zero point energy of the diabatic states, $\hbar \omega/2 \simeq
1800 $ cm$^{-1}$ $ \sim 0.04 D$.
Hence, for smaller $R$ quantum nuclear effects become important. 
The lowest curve in Figure \ref{fig2} shows a ground state potential 
energy surface which has an extremely flat bottom,
as is observed in quantum chemistry calculations for H$_3$O$_2^-$
and H$_5$O$_2^+$ \cite{KowalJCP01}.

The value of $\Delta_1=0.4D \simeq$ 2 eV used here  
for O-H$\cdots$O systems
is estimated from comparisons of the model predictions
with experimental H-bond energies (see Figure \ref{fig4} below) 
and vibrational frequencies (Figures \ref{fig6} and \ref{fig7} ) \cite{note}.
This  is a relatively
large interaction between the diabatic states, and is
comparable to values estimated from quantum chemical calculations
at the MP2 level \cite{SagnellaJCP98}.
The competition between this large energy scale and the
large energies of stretched bonds is key to understanding
H-bond properties.

\begin{figure}[htb] 
\centering 
\includegraphics[width=64mm]{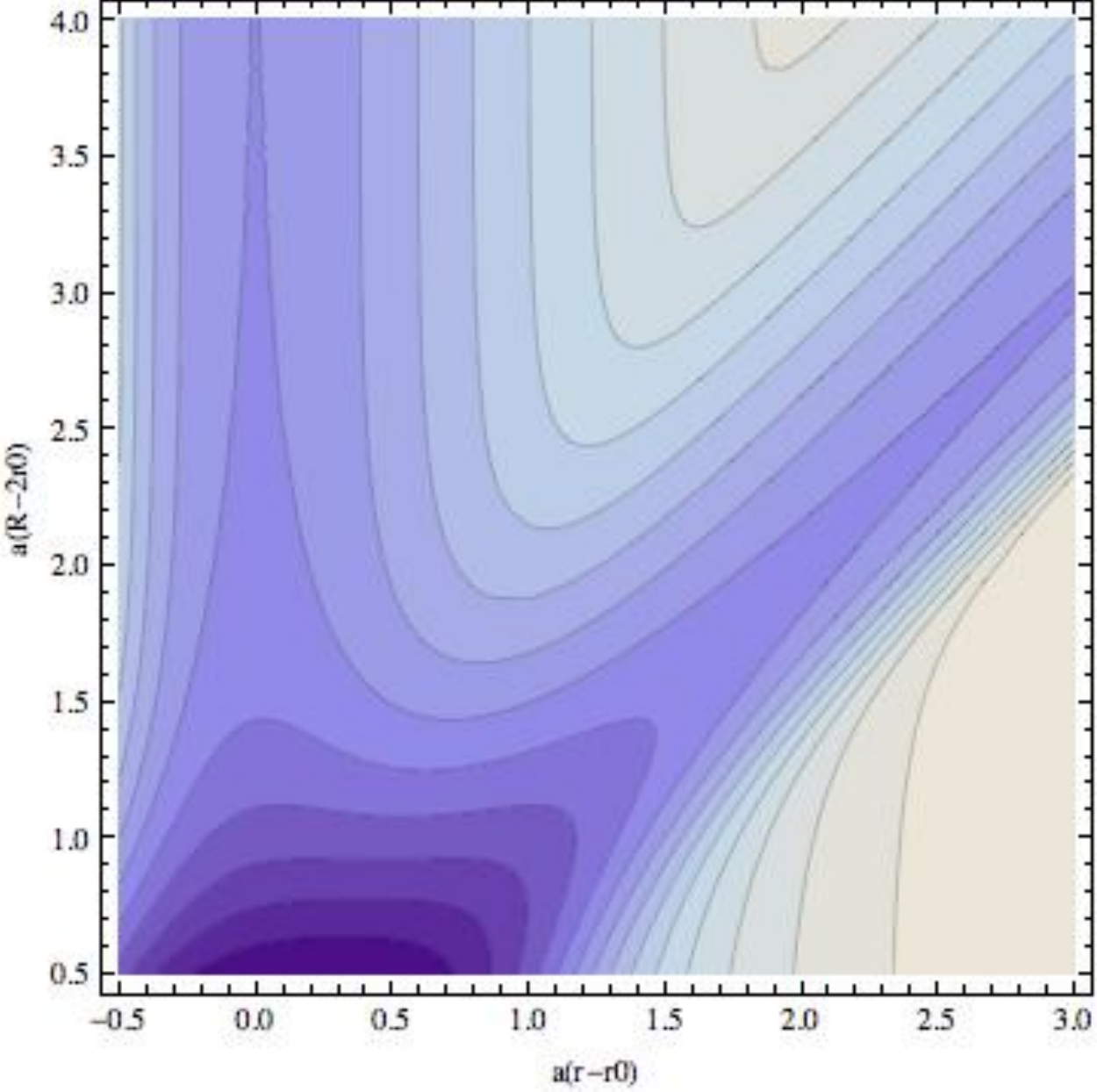}
\caption{(Color online.) 
Contour plot of the ground state potential energy surface 
(for a symmetric donor acceptor-system)
as a function of the D-H bond length $a(r-r_0)$ (horizontal axis)
 and the donor-acceptor distance $a(R-2r_0)$
(vertical axis).
Note that as $R$ varies a wide range
of equilibrium bond lengths $r_m(R)$ are possible.
For this plot $\Delta_1 =0.4 D$ and $b=a$.
The contour spacing is $0.07D$ and darker shades represent lower energies.
}
\label{fig3}
\end{figure}

\begin{figure}[htb] 
\centering  
\includegraphics[width=64mm]{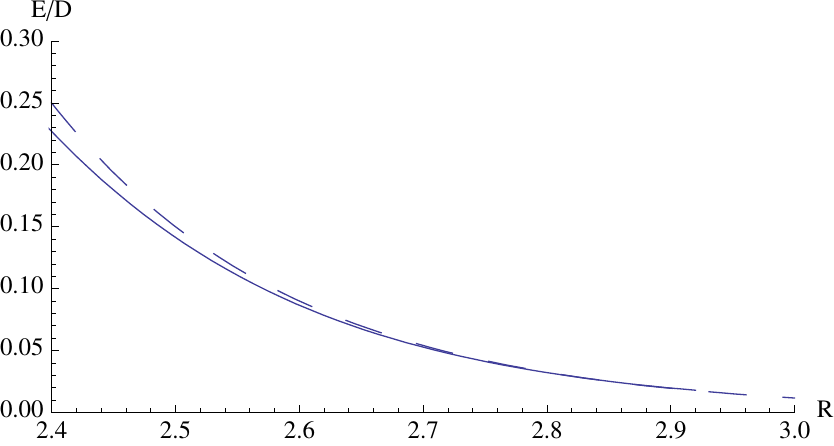}   
\caption{(Color online.) 
Correlation between the
H-bond energy and donor-acceptor distance $R$ (in $\AA$).
Solid line is for the model Hamiltonian
with $b=a=2.2/\AA$, $\Delta_1 =0.4D$.
The dashed curve is an empirical relation,
$E_{HB}=E_{max}\exp(k(R_0-R))$ with $k=5.1$,
$R_0=2.4 \AA$, and $E_{max}=27$ kcal/mol
$=0.25D$ for O-H$\cdots$O bonds\cite{Gilli09}.
}
\label{fig5}
\end{figure}

{\it Insight into the H-bonding puzzle.}
Figures \ref{fig3}, \ref{fig5}, \ref{fig4}
 show that varying $R$  between $2.4 \AA$ and 
$3.0 \AA$ can lead to a wide
range of D-H bond lengths and hydrogen bond energies.
In an actual H-bonded complex the
equilibrium $R$ value will be determined by
the total energy which 
includes contributions from the potential surface shown in Figure \ref{fig3} 
and other interactions not included in this model.
Possible interactions include electrostatic interactions between the
donor and acceptor, steric effects, and van der Waals interactions.
In large  molecules with intramolecular hydrogen
bonding $R$ may actually be determined largely by
the skeletal geometry in which the donor and acceptor 
atom are  imbedded.
Similarly, in water-hydroxyl overlayers on metal surfaces,
the distance $R$ is largely determined by the lattice constant
of the substrate \cite{LiPRL10}.
This provides insight into the origin of the H-bond puzzle because
the identity of the donor and acceptor atoms determines
the Hamiltonian parameters except $R$, whose equilibrium value
will be determined by residual interactions.      

{\it H-bond energies.}
Figure \ref{fig5} shows a favourable comparison
between the calculated binding energies
as a function of $R$ with an empirical relation
for O-H$\cdots$O bonds \cite{Gilli09}.

{\it Bond lengths.}
In the adiabatic limit,  
the equilibrium D-H bond length $r$ 
for a fixed $R$,
is determined by
 the minimum of $E_{-}(r,R)$
 (see Figure \ref{fig4}).
With decreasing $R$ the D-H bond length increases
from its non-interacting value $r_0 \simeq 0.96 \AA$ 
to $R/2$ for a symmetric bond associated with a
barrierless potential well.
Figure \ref{fig4} shows that the bond lengths calculated 
from 
 the minimum of $E_{-}(r,R)$
are significantly less than those observed experimentally for
$R \leq 2.6 \AA$. This difference is attributed to the importance
of quantum nuclear effects which become when  the energy barrier
is comparable to the zero point energy of the D-H stretch vibration.
A signature of such quantum effects are isotope effects.
Indeed this can be seen by comparing the
crystal structure of CrHO$_2$ and CrDO$_2$;
in the former the O-H-O bond appears to be symmetric 
and $R = 2.49 \pm 0.02 \AA$,
whereas the O-D-O bond is asymmetric with an O-D bond length of
$0.96 \pm 0.04 \AA$, with 
$R = 2.55 \pm 0.02 \AA$ \cite{HamiltonAC63}.

\begin{figure}[htb] 
\centering  
\includegraphics[width=64mm]{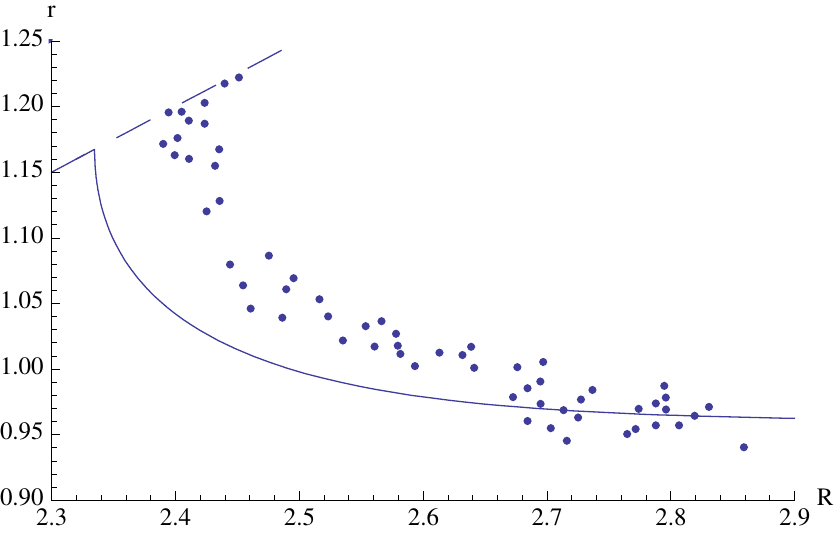}   
\caption{(Color online.) 
Correlation between the 
D-H bond length $r$ and the D-A distance $R$. 
Both lengths are in units of $\AA$.
The solid curves is  the bond length deduced
from the minimum of the adiabatic potential
for $b=a$ and $\Delta(R)=0.4D$.
For the moderate to strong H-bonds which occur for $R < 2.5 \AA$
quantum nuclear motion will significantly increase the D-H bond length
because the energy barrier  becomes comparable to the zero
point energy.
The dots are experimental
data for O-H$\cdots$O bonds in a wide
range of crystal structures and are taken from Figure 6 in \cite{GilliJACS94}.
The dashed line corresponds to symmetric bonds ($r=R/2$).
}
\label{fig4}
\end{figure}

{\it Vibrational frequencies.}
As the distance $R$ decreases there is a significant 
softening of the frequency of the D-H stretch vibration.
In the adiabatic limit this frequency is
given by the curvature at the bottom of the potential
$E_{-}(r,R)$ (see Figure \ref{fig6}).
The latter has been proposed as a measure of the strength of
an H-bond \cite{LiPNAS11}.
Generally, it is expected that when the quantum nuclear
motion is taken into account the actual vibrational 
frequency will be less than adiabatic harmonic frequency.

\begin{figure}[htb] 
\centering
 \includegraphics[width=64mm]{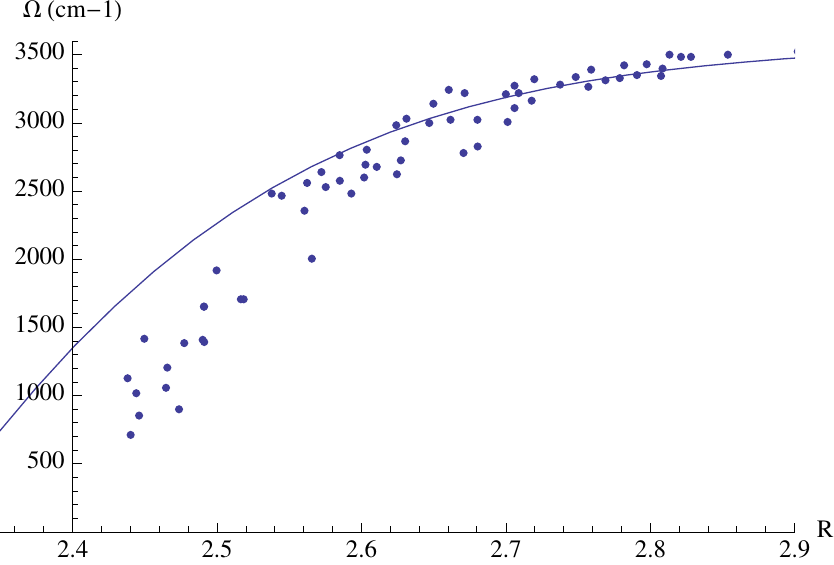}
\caption{(Color online.) 
Softening of the D-H stretch frequency 
$\Omega$ (in cm$^{-1}$)
 with decreasing
donor-acceptor distance $R$ (in $\AA$).
Solid line is the harmonic frequency  for the model Hamiltonian
with $b=a$ and $\Delta_1 =0.4D$.
The dots are experimental data for a wide range
of complexes and are taken from Figure 4 in Ref. \onlinecite{SteinerACIE}.
}
\label{fig6}
\end{figure}

{\it A key prediction of the model}. 
The energy eigenvalue
$E_{+}(r,R)$ describes the potential
energy curve of a low lying excited state
(see Figure \ref{fig2}) which reflects the
quantum mechanical character of the H-bond.
This state would not exist if the H-bond is purely
classical and electrostatic.
This "twin excited state"
 is the analogue of the $^1$B$_{2u}$ state in benzene, of the $^1$B$_2$ state in semibulvalene \cite{Shaik}, and of
the low-lying electronic state in the Creutz-Taube ion
associated with delocalized mixed valence \cite{McKemmish}. 
Hence, for strong hydrogen bond complexes
there should be an electronic excited
state with energy of approximately $2 \Delta \sim 4$ eV 
(corresponding to a wavelength of about 300 nm). The 
 transition dipole moment  is approximately equal to twice
the ground state dipole moment of an isolated OH bond, 
 $2 \vec{d}_{OH} \sim $ 2 Debye,
suggested a significant absorption intensity.
The curve $E_+(r,R)$ has only a single minimum as a function of $r$ and so in this excited state the H-atom will be delocalised between the donor and acceptor.
For strong bonds the curvature of the adiabatic potential of the 
excited state is clearly larger than that of the ground state.
Hence,  the corresponding vibrational frequency will be 
larger than in the ground state, as it is in benzene and semibulvalene \cite{Shaik}.
The intensity of the transition will be reduced by Franck-Condon factors
describing the overlap of the nuclear wave functions in
the ground and excited state.  For weak bonds these overlaps may be small.
Since $\Delta(R)$ decreases with increasing $R$ 
the energy of this excited state will also decrease with
increasing $R$, implying that
exciting to this state will lead to photo-dissociation of the H-bond.
I am   unaware of any experimental investigations
that have looked for this excited state in the UV.

{\it Asymmetric bonds.}
We now consider the case where the proton affinity (PA) of the donor
($D_D$) and of the
acceptor ($D_A$) are unequal, i.e., $\epsilon \equiv D_D - D_A \neq 0$. 
Gilli and Gilli \cite{Gilli09}
 have noted the "PA, $pK_a$ equalisation principle": the 
strongest hydrogen bonds occur when the donor and acceptor have the same
proton affinity.
This occurs naturally in our two-state Hamiltonian. For example, for
very short D-A distances and 
large $\Delta$, the H-bond energy is approximately, $\epsilon - \sqrt{\epsilon^2 + \Delta^2}$,
which has its largest value for $\epsilon=0$. 
Figure \ref{fig7} shows the softening of the D-H stretch frequency as
a function of $\epsilon$, 
compared  to experimental data \cite{RoscioliScience07}.

\begin{figure}[htb] 
\centering 
\includegraphics[width=64mm]{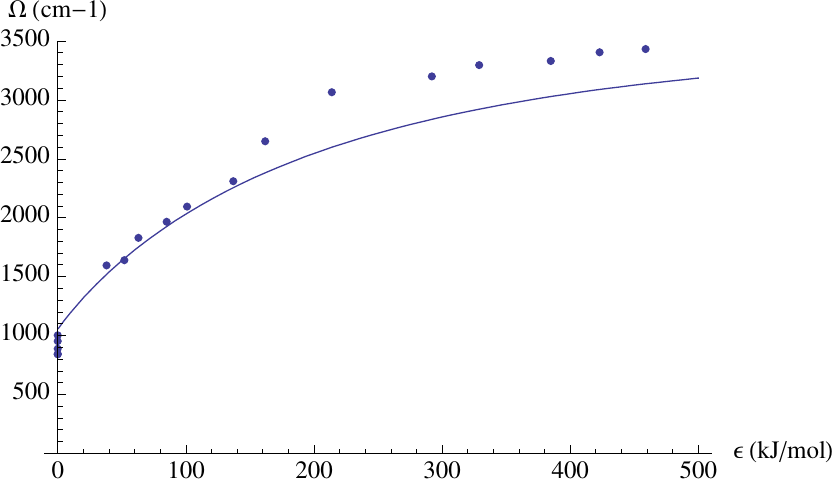}       
\caption{(Color online.) 
Softening of the D-H stretch frequency $\Omega$ 
(in units of cm$^{-1}$)
with decreasing
difference between the proton affinity of the acceptor and the donor
($\epsilon$ in units of kJ/mol).
The solid line is the harmonic frequency of the model Hamiltonian with
$R=2r_0+1/a \simeq 2.4 \AA$,  $\Delta_1 =0.4D_A$, $D_A=120$ kcal/mol.
The dots are experimental data \cite{RoscioliScience07}.
}
\label{fig7}
\end{figure}


{\it Quantum and isotope effects.}
The above discussion treated the nuclear degrees of freedom classically,
but noted that quantum nuclear effects may have a significant
effect on equilibrium bond lengths, particular for strong H-bonds.
Significant isotope effects are observed experimentally \cite{SteinerACIE}
and arise due to the
 zero point motion of the hydrogen atom.
Furthermore, quantum nuclear effects may play an important role in 
water \cite{MorronePRL08}, in water-hydroxyl overlayers on
metal surfaces \cite{LiPRL10}, and in some 
proton transfer reactions in enzymes \cite{BothmaNJP10}.
For strong bonds where the potential is very anharmonic quantum nuclear effects
will also be important \cite{LiPNAS11}.
The model Hamiltonian
provides a natural means to describe these effects if the 
the hydrogen atom co-ordinate $r$ is treated quantum mechanically.
The harmonic limit corresponds to a spin-boson model which has an
analytical solution in terms of continued fractions \cite{PaganelliJPCM06}.
The fully quantum Morse potential has an exact analytical solution and 
an algebraic representation in terms of creation and annihilation
operators \cite{Levine}.  
Hence, an algebraic treatment of the quantum version of the model Hamiltonian
may also be possible, because the off-diagonal terms are
independent of $r$. 



In conclusion, the relatively simple effective Hamiltonian 
considered here provides a unified picture of a range of phenomena
associated with hydrogen bonding.
Futhermore, it predicts an excited state which should
lead to photo-disassociation of an H-bonded complex.
Future work could consider
non-linear bonds and the associated vibrational bending modes,
and correlations between $R$ and vibrational absorption intensities
and NMR chemical shifts.
It would be worthwhile to provide a more
rigorous justification of the diabatic state Hamiltonian from quantum
chemistry. This may be done in a similar systematic manner as      
has been done for the excited states of the chromophore
of the Green Fluorescent Protein \cite{OlsenJCP09}.

\begin{acknowledgments}
I thank N. Hush, S. Olsen and J. Reimers for providing
many insights about diabatic states.
Discussions with X. Huang are also acknowledged.                   
I thank X. Li, L. McKemmish, J. Reimers,  and S. Shaik for
helpful comments on a draft manuscript.
Financial support was received from an
  Australian Research Council Discovery Project grant (DP0877875).
\end{acknowledgments}


\begin{thebibliography}{39}
\bibitem{note}
It can be shown that the condition for no energy barrier
in the ground state is
\begin{equation}
\frac{\Delta(R)}{D}
 \geq \frac{(1-\exp(-\mu))^2}{1-\frac{1}{2}\exp(\mu)}
\end{equation}
where $\mu \equiv a(R/2-r_0)$ \cite{note}.    
This leads to a symmetric D-H-A bond in the adiabatic
limit where the nuclear degrees of freedom are treated classically. 
 Empirical correlations \cite{Gilli09} suggest
that for O-H-O complexes symmetric bonds occur when $R \simeq 2.4 \AA$.
Using $a \simeq 2.2/\AA, r_0=0.96 \AA$ gives
 $\mu = 0.52$
the above expression gives $\Delta(R) \simeq D$.
For short bonds (i.e., $ \mu \ll 1$) the right hand side
the above expression 
reduces to $ 2\mu^2 $ (harmonic limit) which for
$\mu = 0.52$ gives $\Delta \simeq 0.6D$, significantly different
from the value above,      showing
 the importance of anharmonic effects.
\end{thebibliography}
\end{document}